\documentclass{ws-ijbc}
\usepackage{sidecap}
\sidecaptionvpos{figure}{c}

\usepackage{amsmath}
\usepackage{amssymb}
\usepackage{bm}
\usepackage{epsfig}

\newcommand{\abs}[1]{\vert #1 \vert}
\newcommand{\ie}{\textit{i.e.~}}
\newcommand{\eg}{\textit{e.g.~}}
\newcommand{\etal}{\textit{et al.~}}

\begin{document}
\catchline{}{}{}{}{} 

\markboth{J.D. Bodyfelt, T.V. Laptyeva, G. Gligoric, D.O. Krimer, Ch. Skokos, \& S. Flach}
{Wave interactions in localizing media - a coin with many faces}
\title{(This paper is for the Special Issue edited by \\ Prof. Gregoire Nicolis , Prof. Marko Robnik, Dr. Vassilis Rothos and Dr. Haris Skokos) \\ Wave interactions in localizing media - a coin with many faces}
\author{J.D. Bodyfelt}
\address{Max-Planck-Institut f\"{u}r Physik komplexer Systeme, N\"{o}thnitzer Stra\ss e 38, D-01187 Dresden, Germany}
\author{T.V. Laptyeva}
\address{Max-Planck-Institut f\"{u}r Physik komplexer Systeme, N\"{o}thnitzer Stra\ss e 38, D-01187 Dresden, Germany}
\author{G. Gligoric}
\address{Max-Planck-Institut f\"{u}r Physik komplexer Systeme, N\"{o}thnitzer Stra\ss e 38, D-01187 Dresden, Germany}
\author{D.O. Krimer}
\address{Max-Planck-Institut f\"{u}r Physik komplexer Systeme, N\"{o}thnitzer Stra\ss e 38, D-01187 Dresden, Germany \\
Theoretische Physik, Universit\"{a}t T\"{u}bingen, Auf der Morgenstelle 14, 72076  T\"{u}bingen, Germany}
\author{Ch. Skokos}
\address{Max-Planck-Institut f\"{u}r Physik komplexer Systeme, N\"{o}thnitzer Stra\ss e 38, D-01187 Dresden, Germany \\
Center for Research and Applications of Nonlinear Systems, University of Patras, GR-26500 Patras, Greece}
\author{S. Flach}
\address{Max-Planck-Institut f\"{u}r Physik komplexer Systeme, N\"{o}thnitzer Stra\ss e 38, D-01187 Dresden, Germany}

\maketitle
\begin{history}
\received{(to be inserted by publisher)}
\end{history}
\begin{abstract}
A variety of heterogeneous potentials are capable of localizing linear non-interacting waves. In this work, we review different examples
of heterogeneous localizing potentials which were realized in experiments. We then discuss the impact of nonlinearity induced by wave
interactions, in particular its destructive effect on the localizing properties of the heterogeneous potentials.
\end{abstract}
%
From communications amongst ourselves to sensing and imaging our surroundings, wave transmissions through media play an integral role in our 
modern lives. Regardless of their physical origin or scale, most wave systems exhibit phenomena such as scattering, reflection, refraction, 
and interference. This commonality thus provides a framework to discuss universal features. Transmission media, modeled by mathematical 
potentials, are often idealized with the characteristics of integrability, hermiticity (\ie closed systems), linearity, and homogeneity. 
In-situ applications rarely however are at this level of perfection - even in a laboratory, such situations require colossal experimental 
control and accuracy to achieve. Besides, these small differences from perfection may lead to novel and utilitarian phenomena.

One such small difference is heterogeneity of the medium. This heterogeneity can arise in several different ways, but throughout, attenuation of 
the wave transmission is possible - a phenomenon called \textit{localization}. The first heterogeneity to be discussed is the presence of 
randomly distributed impurities, leading to an exponential localization in the eigenstates - called \textit{Anderson localization} (AL). 
In Sec.~\ref{sec:localization}, this AL is described in greater detail. The second form of heterogeneity arises from the 
presence of a DC-biased electric field, creating an equidistant spectrum and an inverse factorial eigenstate attenuation - 
called \textit{Wannier-Stark localization}. This case is detailed within Sec.~\ref{sec:lin_stark}. The third case, covered in Sec.~\ref{sec:quasi}, 
comes from potentials that arranged in correlated fashions. In these types of heterogeneity, a tuning parameter may allow transitions from 
localized states to those that are unattenuated, with critical states at the transition point. The last form of heterogeneity to be discussed 
arises from a specially designed potential, in which localization occurs not spatially, but within a momentum space. This 
\textit{dynamical localization} will be presented in Sec.~\ref{sec:dynamic}. 

In addition to heterogeneity, a second deviation from ideal transmission media is due to wave interaction, resulting in nonlinear dependencies on
wave amplitudes. Examples of such nonlinear potentials are prolific, and include: the AC Kerr effect in optical media, 
atom-atom scattering in boson condensates, screened Coulomb interactions in electrons, and acoustic Langmuir waves in cold plasmas. 
This nonlinearity can additionally be coupled to heterogeneous localizing potentials, drastically altering the resulting localization. 
Within Sec.~\ref{sec:nonlinear}, the effect of nonlinearity on the dynamics of packets in localizing media is discussed, with individual 
subsections corresponding to the four different heterogeneities. 

In short, we address what happens when first, the linear waves yield zero conductivity (localized) in heterogeneous media, and then wave 
interactions are added. Will an insulator change into a conductor, or will localization remain? This is the main question 
in the interplay between heterogeneity and nonlinearity, and helps define how these two different ingredients collaborate. 
\section{Linear Waves in Localizing Media}
\subsection{Disorder - Anderson Localization}\label{sec:localization}
A fundamental problem of condensed matter physics was (and still remains) the study of conductivity of electrons in solids. Since in
an infinite perfect crystal, electrons can propagate ballistically, a natural question is raised: what happens in a more realistic situation when
there is disorder in the crystal due to impurities or defects? Will the increase of the degree of disorder lead to a decrease of conductivity, or not?  
These questions were first answered in a seminal paper by P.W. Anderson~\cite{anderson_absence_1958}, where it was shown that for large enough strengths 
of disorder the diffusive motion of the electron will come to a halt.  In particular, Anderson studied an unperturbed lattice of uncoupled sites, where 
the perturbation was considered to be the coupling between them, and randomness was introduced in the on-site energies. For this model he showed that 
for a large degree of randomness, the transmission of a wave decays exponentially with the length of the lattice.

This absence of wave diffusion in disordered mediums is nowadays called \textbf{Anderson localization}, and is a general wave phenomenon 
that applies to the transport of different types of classical or quantum waves, like electromagnetic, acoustic and spin waves. Its origin 
is the wave interference between multiple scattering paths; \ie the introduction of randomness can drastically disturb the constructive 
interference, leading to the halting of waves. Anderson localization plays an important role in several physical phenomena. For example, 
the localization of electrons has dramatic consequences for the conductivity of materials, since the medium no longer behaves like a metal,
but becomes an insulator when the strength of disorder exceeds a certain threshold. This transition is often referred as the metal-insulator 
transition (MIT).

Often theoretical and numerical approaches of localization start with the Anderson model:  a standard tight-binding (\ie 
{nearest-neighbor} hopping) with an on-site potential disorder. This can be represented in one dimension by a time-dependent Schr\"{o}dinger 
model
\begin{equation}
i \frac{\partial \psi_l}{\partial t} = \epsilon_l \psi_l - \psi_{l+1}- \psi_{l-1} \label{eq:Anderson}
\end{equation}
Here $\{\epsilon_l\}$ are the random on-site energies, which are drawn from an uncorrelated uniform distribution in $[-W/2, W/2]$, 
where $W$ parametrizes the disorder strength. The $\psi_l$ is the complex wavefunction associated with lattice site $l$. 
Using the substitution $\psi_l = A_l \exp(-i \lambda t)$ yields a time-independent form
\begin{equation}
\lambda A_l=\epsilon_l A_l - A_{l+1}- A_{l-1},
\label{eq:Schrodinger}
\end{equation} 
The solution consists of both a set of eigenvectors called the \textbf{normal modes} (NMs), $A_l^\nu$, (normalized as 
$\sum_{l} (A_l^\nu)^2=1\,$), and also a set of eigenvalues called the normal frequencies, $\lambda_{\nu} \in [-W/2-2, W/2 + 2]$
which exist in a spectral band of width $\Delta = 4 + W$. The eigenvectors are exponentially localized, meaning that their 
\textit{asymptotic behavior} can be described by an exponential decay
\begin{equation}
 \abs{A_l^\nu} \sim e^{-l/\xi(\lambda_\nu)},
\label{eq:decay}
\end{equation} 
where $\xi(\lambda_\nu)$ is a characteristic energy-dependent length, called the \textit{localization length}. Naturally, $\xi \rightarrow \infty$ 
corresponds to an extended eigenstate. Several approaches have been developed for the evaluation of $\xi$, such as: the transfer matrix method, 
schemes based on the transport properties of the lattice, and perturbative techniques. For more information on such approaches, the reader is referred 
to \cite{kramer_localization_1993} and references therein. In general, these approaches approximate the localization length as $\xi(\lambda_\nu) 
\approx 96/W^2$ for weak disorder strengths, $W\leq 4$. On average, the localization volume (\ie spatial extent) $V$ of the NM is on order of 
$3.3\xi(0)$ for these weak disorder strengths, and tends to unity in the limit of strong disorder.

In real experiments, measurements of transmission and conductivity are mainly performed, so the need for a connection between the
conductivity and the spectrum becomes apparent. The basic approach towards the fulfillment of this goal was the establishment of a connection
between the conductivity and the sensitivity to changes of the boundary conditions of the eigenvalues of the Hamiltonian of a finite (but 
very large) system \cite{edwards_numerical_1972}. The sensitivity to the boundary conditions turned out to be conceptually important for 
the formulation of a scaling theory for localization \cite{abrahams_scaling_1979}. The main hypothesis of this single-parameter scaling 
theory is that close to the transition between localized and extended states, there should be only one scaling variable which should 
depend on the conductivity for the metallic behavior, and the localization length for the insulating behavior. This single parameter 
turned out to be a dimensionless conductance (often called \textit{Thouless conductance} or \textit{Thouless number}) defined as
\begin{equation}
 g(N)=\frac{\delta E}{\Delta E},
\label{eq:Tc}
\end{equation} 
where $\delta E$ is the average energy shift of eigenvalues of a finite system of size $N$ due to the change in the boundary
conditions, and $\Delta E$ is the average spacing of the eigenvalues. For localized states and large $N$, $\delta E$ becomes very small 
and $g(N)$ exponentially vanishes. In the metallic regime the boundary conditions always influence the energy levels, even in the limiting 
case of infinite systems. The introduction of the Thouless conductance led to the formulation of a simple criterion for the occurrence of Anderson
localization: $g(N)<1$. In one and two-dimensional random media this criterion can be reached for any degree of disorder by just increasing
the size of the medium, while in higher dimensions a critical threshold exists.
 
The experimental observation of Anderson localization is not easy, for example due to the electron-electron interactions in cases of electron
localization, and the difficult discrimination between localization and absorption in experiments of photon localization. Nevertheless,
nowadays the observation of Anderson localization has been reported in several experiments, a few of which we quote here. In
\cite{wiersma_localization_1997,stoerzer_observation_2006} localization of light in three-dimensional random media was reported. 
Anderson localization has also been observed in experiments of transverse localization of light for two \cite{schwartz_transport_2007}
and one \cite{lahini_anderson_2008} dimensional photonic lattices. Anderson localization has also been observed in experiments of 
localization of a Bose-Einstein condensate in an one-dimensional optical potential \cite{billy_direct_2008,roati_anderson_2008}, 
and of elastic waves in a three-dimensional disordered medium \cite{hu_localization_2008}. In addition, the observation of the 
MIT in a three-dimensional model with atomic matter waves has been reported \cite{chabe_experimental_2008}.
\subsection{Wannier-Stark Ladder - Bloch Oscillations}\label{sec:lin_stark}
Another intriguing class of problems appears when replacing the disorder potential $\epsilon_{l}$ by a DC field $\epsilon_{l}=l E$ 
($E$ denotes the strength of the field). As an example, one can mention the textbook solid state problem of an electron in a periodic potential 
with an additional electric field (see e.g. \cite{tsu_field_1993}) which leads to investigations of Bloch oscillations \cite{wannier_wave_1960} 
and Landau-Zener tunneling \cite{landau_theory_1932,zener_non-adiabatic_1932,liu_theory_2002} in different physical systems. Nowadays, such 
effects are experimentally observed employing optical waves in photonic lattices \cite{pertsch_optical_1999,morandotti_experimental_1999} and 
ultracold atoms in optical lattices \cite{anderson_macroscopic_1988,gustavsson_control_2008,morsch_bloch_2001}. If the well depth of the 
periodic potential is large enough, Landau-Zener tunneling is suppressed, the problem is discretized (Wannier-Stark ladder) and the resulting 
eigenvalue problem is explicitly solved in terms of the localized eigenmodes of the system (see \eg \cite{fukuyama_tightly_1973}). Let us consider a 
discrete linear Schr\"{o}dinger equation, as in Eq.(\ref{eq:Anderson}) but now with a DC bias $E$:
\begin{equation}
i\dot\psi_l=l E \psi_l -\psi_{l+1}-\psi_{l-1}
\label{eq:Lin_Stark}
\end{equation}
Note this is the governing equation \eg for dilute Bose-Einstein condensate dynamics in a deep and biased optical potential, whereby 
$|\psi_l(t)|^2$ has a meaning of BEC density in the $l$-th potential well. Here also $\psi_l$ is a site's complex amplitude, and 
the same substitution as in Eqs.(\ref{eq:Anderson},\ref{eq:Schrodinger}) can be made to arrive at the eigenvalue problem
\begin{equation}
\lambda A_l=l E A_l -A_{l+1}-A_{l-1} \label{eq:Lin_Stark_Eigen}
\end{equation}
In the case of an infinite lattice, this yields eigenvalues of $\lambda_\nu=E\nu$ (with integer $\nu$). These eigenvalues form an equidistant 
spectrum which extends over the whole real axis - the Wannier-Stark ladder. The corresponding normal modes obey the generalized translational 
invariance $A_{l+\mu}^{\nu+\mu}=A_l^\nu$ \cite{wannier_wave_1960} and are given by the Bessel functions of the first kind, $A_l^{(0)}=J_l(2/E)$. 
All normal modes are spatially localized with an asymptotic decay of
\begin{equation*}
|A_{l \rightarrow \infty}^{(0)}| \rightarrow \left(1/E\right)^l/l!
\end{equation*}
The eigenstates are thus more strongly localized as compared to the disordered case, where the eigenstates decay exponentially. Remarkably, 
various observables exhibit temporal periodic motion (Bloch oscillations) with the period $T_B = 2\pi / E$.

\begin{SCfigure}[10][htb]
\centering
\includegraphics[width=0.45\textwidth, keepaspectratio,clip]{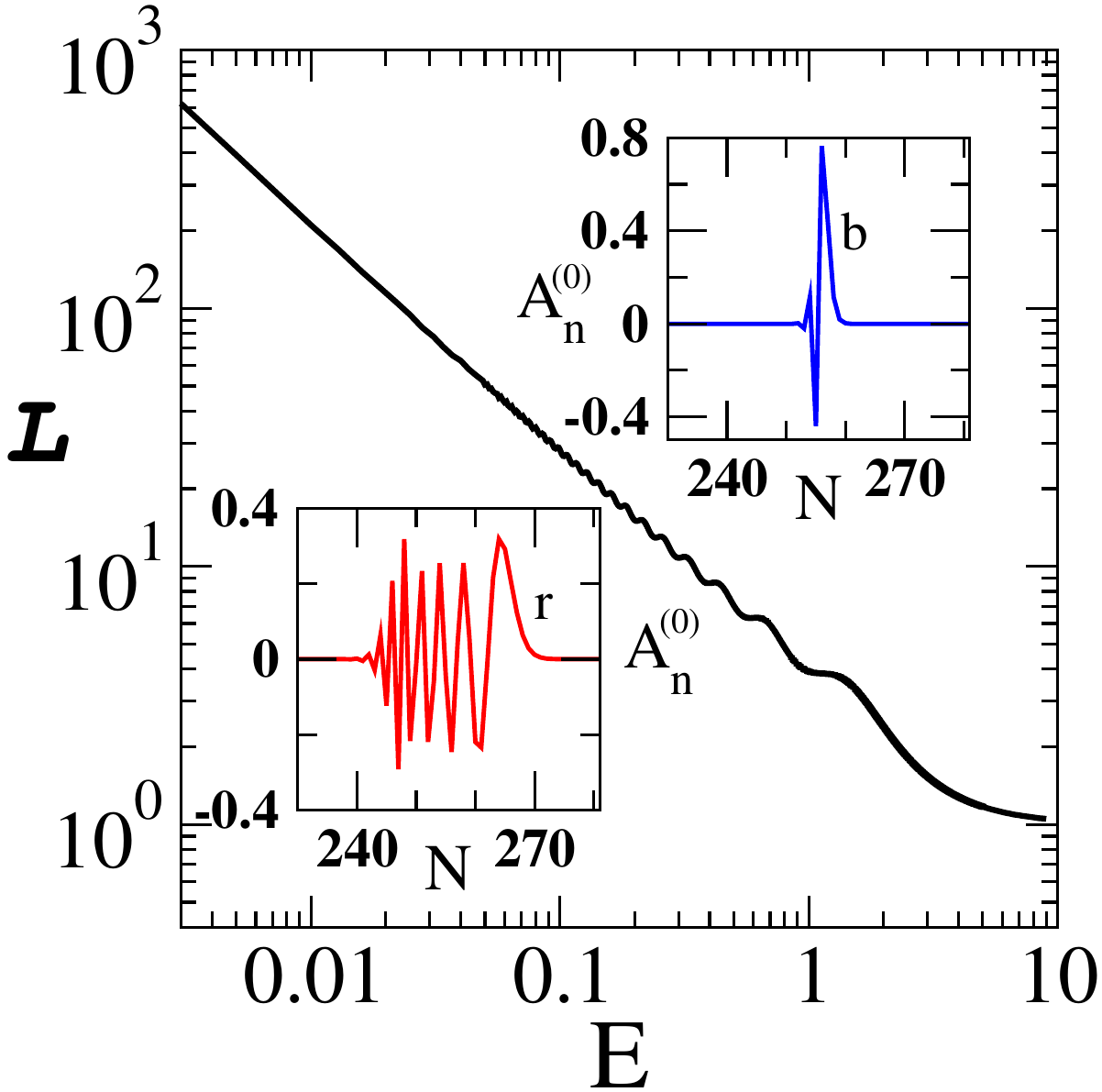}  
\caption{(Color online) Localization volume $\mathcal{L}$ of the eigenfunction $A_n^{(0)}$ versus $E$.  
Insets: Explicit form of the eigenfunction across the chain for two values $E=2$ and  $E=0.2$ [(b), blue; (r), red].
Figure adapted from \cite{krimer_delocalization_2009}.}
\label{fig:Lin_Stark}
\end{SCfigure}

The localization volume of an eigenstate is determined via $\mathcal{L}=1/\sum_l |A_l^{(0)}|^4$ \cite{krimer_delocalization_2009}. It characterizes 
the spatial extent of the eigenstate as a function of $E$. In Fig.~\ref{fig:Lin_Stark}, the localization volume of an eigenstate is shown as a 
function of $E$. The asymptotic behavior was found to be $\mathcal{L}\propto -[E\cdot\ln E]^{-1}$ for $E\rightarrow 0$, and $\mathcal{L} \rightarrow 1$ 
for $E\rightarrow \infty$ \cite{krimer_delocalization_2009}.
\subsection{Aubry-Andr\'{e} Chains - An Example of Quasiperiodicity}\label{sec:quasi}
A periodic lattice gives a translational symmetry. A simple way to destroy the symmetry is to introduce a secondary periodic lattice of a different and 
incommensurate frequency. This idea has gained much attention in solid-state via \textit{quasicrystals}~\cite{levine_quasicrystals_1984,
trebin_quasicrystals_2003,vekilov_quasicrystals_2010}. The idea of incommensuration also extends into optics via sequenced (\eg Fibonacci, Thue-Morse, 
Rudin-Shapiro) potentials~\cite{macia_electrons_2000, albuquerque_theory_2003}. More so, it has recently been focused upon in ultracold atomic physics, 
in terms of \textit{bichromatic} lattices~\cite{guidoni_quasiperiodic_1997, modugno_exponential_2009, modugno_anderson_2010, albert_localization_2010}. 
 
Regardless of the subfield, for large lattices the dynamics is described in a tight-binding form - as in Eqs.(\ref{eq:Anderson},\ref{eq:Lin_Stark}) -
and goes by the moniker of \textit{Aubry-Andr\'{e} model}
\begin{equation}
i \frac{\partial \psi_l}{\partial t} =  \zeta \cos(2 \pi \alpha l) \cdot \psi_l -  \psi_{l+1} - \psi_{l-1}
\label{eq:harper}
\end{equation}
in which the parameter $\alpha$ dictates the commensurability ratio between the two different frequencies. The parameter $\zeta$ dictates a relative lattice 
strength, much as $W$ for the disordered Anderson model. As in Eqs.(\ref{eq:Schrodinger},\ref{eq:Lin_Stark_Eigen}), a substitution is made to turn 
Eq.(\ref{eq:harper}) into an eigenvalue problem:
\begin{equation}
\lambda A_l =  \zeta \cos(2 \pi \alpha l) \cdot A_l -  A_{l+1} - A_{l-1}
\label{eq:harper_eigen}
\end{equation}

Originally this model was introduced by Harper~\cite{harper_single_1955, harper_general_1955}  (hence Eq.(\ref{eq:harper}) is also synonymously the 
\textit{Harper model}) to describe a low-temperature two-dimensional electron gas in a high magnetic field, in which the parameter $\alpha$ 
describes incommensurability between the quantum of magnetic flux and the lattice cell. 
It is commonplace to make the lattice as largely incommensurate 
as possible for studies in the Aubry-Andr\'{e} model; the inverse golden mean is often used $\alpha = \frac{\sqrt{5}-1}{2}$. We shall henceforth keep with 
this standard convention.

\begin{SCfigure}[10][htb]
\centering
\includegraphics[width=0.4\textwidth, keepaspectratio,clip]{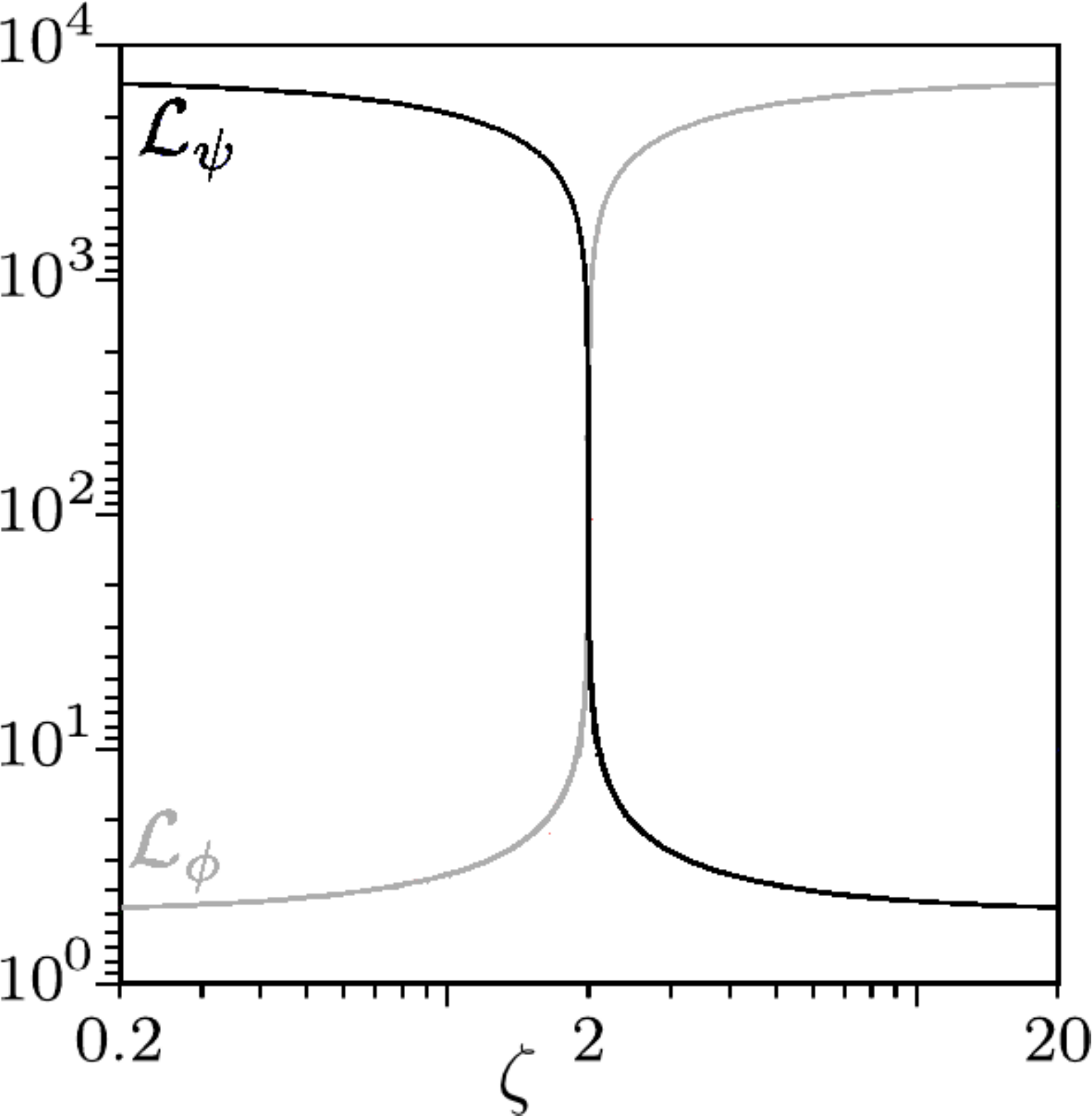}  
\caption{The self-duality of the Aubry-Andr\'{e} model: For lattice size of 10946, as a function of the parameter $\zeta$ is shown the average 
localization volumes in both real space ($\mathcal{L}_\psi$, black) and momentum space ($\mathcal{L}_\phi$, gray). For small $\zeta < 2$, 
the eigenstates are extended in real space ($\mathcal{L}_\psi \gg 1$) and localized in momentum space ($\mathcal{L}_\psi \sim 1$). At 
$\zeta=2$ the exchange is seen, \ie for $\zeta > 2$ we see the eigenstates localized in real space and extended in momentum space. 
Figure adapted from those shown in \cite{aulbach_phase-space_2004}.}
\label{fig:harper_ipr}
\end{SCfigure}

Using the Fourier form $\psi_l = \sum_k e^{2 \pi i \alpha k l} \phi_k$, Eq.(\ref{eq:harper}) transforms into the quasi-momentum basis of $\{\phi_k\}$ 
\begin{equation}
i \frac{\partial \phi_k}{\partial t} =   2\cos(2\pi\alpha k)\cdot\phi_k - \frac{\zeta}{2} \phi_{k+1} -\frac{\zeta}{2} \phi_{k-1} 
\label{eq:harper_trans}
\end{equation}
Note that Eqs.(\ref{eq:harper},\ref{eq:harper_trans}) are \textbf{not} identical, as seen in the location of the $\zeta$ parameter. 
Eq.(\ref{eq:harper}) dictates dynamics in a position representation and $\zeta$ occurs on the on-site energy terms, while Eq.(\ref{eq:harper_trans}) 
dictates dynamics in a momentum representation, and $\zeta$ appears in the kinetic coupling of the momentum modes. Even though the two equations are 
strictly different, under exchange of $\zeta$ the two equations are equivalent in form - a property know as \textit{self-duality}, first derived by Aubry 
and Andr\'{e} \cite{aubry_analyticity_1980}. Both equations easily can be seen to be equivalent without ANY parameter exchange if $\zeta=2$. 
This self-dual symmetry is nicely observed in the eigenstates. 
For the critical value of $\zeta=2$, the two representations have identical localization lengths. Additionally, the localization volumes can
be probed as in the prior section, using either $\mathcal{L}_\psi=1/\sum_l |\psi_l|^4$ or $\mathcal{L}_\phi=1/\sum_k |\phi_k|^4$. This is shown 
in Fig.~\ref{fig:harper_ipr}. In the figure, for small $\zeta$ states localized in position space are extended in momentum space. As $\zeta$ sweeps 
across $\zeta=2$, the strong transition from localized to extended is seen for $\mathcal{L}_\psi$ (and vice-versa for $\mathcal{L}_\phi$).

The value $\zeta = 2$ is thus tied strongly to criticality and fractality of eigenstates, whose existence in quasiperiodic models has been 
quite an active hotbed. In terms of the Aubry-Andr\'{e} model, this has focused on changes in the density of states~\cite{soukoulis_localization_1982}, 
density-density correlations~\cite{boers_mobility_2007,li_exploring_2010}, Husimi/Wigner distributions~\cite{ingold_delocalization_2002, 
aulbach_phase-space_2004}, and spreading of density moments~\cite{hu_disturbance_2000, hufnagel_superballistic_2001, diener_transition_2001}. 
The Aubry-Andr\'{e} model has also seen modification by another parameter $\nu$, introduced in Eq.(\ref{eq:harper}) as $\cos(2\pi\alpha l) \mapsto 
\cos(2\pi\alpha n^{\nu})$, in order to further probe and control the mobility edge \cite{griniasty_localization_1988,das_sarma_mobility_1988, 
varga_power-law_1992}. Study of the Aubry-Andr\'{e} model continues onward, beyond simple tight-binding formalisms
\cite{johansson_self-dual_1991,biddle_localization_2009}.
\subsection{Quantum Kicked Rotor - Localization in Momentum Space}\label{sec:dynamic}
There is a growing interest in the study of quantum systems with time-dependent Hamiltonians. An important motivation in this area is better 
understanding of the quantum dynamics within simple systems whose classical counterparts exhibit chaotic behavior. One of the relatively simple 
models to study the quantum dynamics is a quantum kicked rotor. This model was introduced in \cite{casati_stochastic_1979} as a quantum analog of 
\textit{the standard mapping} \cite{chirikov_universal_1979}, which is the basic model of dynamical chaos in the classical limit. In spite of this 
formal analogy, the dynamical chaos in the quantum kicked rotor exhibits some specific features closely related with the quantum nature of the 
underlying model. Namely, in the classical case the motion depends on a single parameter, $K$, the dimensionless strength of kick. For each value of $K$
the motion can be quasi-periodic, chaotic, or accelerating, depending on the initial conditions. At small $K$ the chaotic regions are isolated and separated by the
Kolmogorov-Arnold-Moser (KAM) trajectories; consequently the motion is bounded. For $K=K_c\approx 0.97146$ the last of these trajectories disappears 
and diffusion in the momentum space takes place. On the other hand, in the quantum case the energy remains bounded and does not increase with time even 
for $K>K_c$. In other words, the quantum suppression of classical diffusion in the phase space has taken place in the model of quantum kicked rotor
\cite{izrailev_simple_1990}. The quantum localization of classical chaotic diffusion is sometimes called \textit{dynamical localization}. This phenomenon is 
in many aspects analogous to Anderson localization in the models with disorder \cite{fishman_chaos_1982}. However, in the case of quantum chaos there is no 
randomness and transient diffusion in the corresponding classical system. In other words, the dynamical localization in a quantum kicked rotor occurs in a 
completely deterministic system. In addition, in the cases where the period of kicks equals to an integer multiple of the natural period of rotor, quantum 
resonances and ballistic diffusion occur.

The quantum kicked rotor is described by the Schr\"{o}dinger equation:
\begin{equation}
i\frac{\partial \psi}{\partial t} = H \psi = -\frac{1}{2}\frac{\partial^{2} \psi}{\partial \theta^{2}}+k\cos\left(\theta\right)\psi 
\sum^{+\infty}_{m=-\infty} \delta \left(t-mT\right) \,\,.\label{this01}
\end{equation}
Here, $\theta$ and $-i\partial/\partial \theta$ are the position and the conjugated momentum operators of the rotor. All quantities are in units 
of $\hbar=1$, and the motion is considered on a ring with periodic boundary conditions $\psi\left(\theta+2\pi\right)=\psi\left(\theta\right)$. The
parameter $k$ is the kick strength and $T$ is the period between kicks. The evolution operator over one period $T$ is given by
\begin{equation}
\hat{U}=\exp\left(-i\frac{T}{2}\frac{\partial^{2}}{\partial \theta^{2}}\right)\exp\left(-ik\cos\left(\theta\right)\right)\,\,
.\label{this11}
\end{equation}
The solution $\psi\left(\theta,t\right)$ of Eq.(\ref{this01}) can be expanded in the basis of the angular momentum eigenfunctions in a form
\begin{equation}
\psi\left(\theta,t\right)=\frac{1}{\sqrt{2\pi}}\sum^{\infty}_{n=-\infty}A_{n}\left(t\right)\exp\left(i n \theta\right)\,\,
,\label{this21}
\end{equation}
where the coefficients $A_{n}\left(t\right)$ are the Fourier coefficients of the time-dependent wave function $\psi\left(\theta,t\right)$. 
As a result of the action of the evolution operator, Eq.(\ref{this11}), on the wave function $\psi\left(\theta,t\right)$ over one period $T$, 
the following mapping of the Fourier coefficients $A_{n}$ is obtained
\begin{equation}
A_{n}\left(t+T\right)=\sum_{m}\left(-i\right)^{n-m}J_{n-m}\left(k\right)A_{m}\left(t\right)\exp\left(-i\frac{1}{2}Tm^{2}\right)\,\,
,\label{this31}
\end{equation}
where $J_{n-m}\left(k\right)$ is the Bessel function of the first order \cite{casati_stochastic_1979}. It is found from Eq.(\ref{this31}) that -
opposite to the classical model where one parameter determines the system behavior - in the quantum model, behavior depends on two parameters: 
$k$ and $T$ \cite{izrailev_simple_1990}. The perturbation strength $k$ gives the effective number of unperturbed states covered by one kick, 
and $T$ is ratio between the period of kicks $T$ and the natural period of rotor, set to one in this case. When the ratio between these 
two periods is rational (\ie $T$ is a rational number), the rotor energy $E\left(t\right)=\sum_{n}\left|A_{n}\left(t\right)\right|^{2}n^{2}/2$, 
grows ballistically in time as $t^2$, at variance to the classical case. This phenomenon is called quantum resonance (being caused by pure quantum 
interference effects) and has no relation to the classical behavior \cite{casati_stochastic_1979}. On the other hand, for irrational $T$, suppression 
of the energy diffusion occurs and the spreading of the wave packet stops. Eq.(\ref{this31}) can be also treated as an eigenvalue problem 
\begin{equation}
\lambda_\nu A_l^\nu = \sum_{m}\left(-i\right)^{n-m}J_{n-m}\left(k\right)\exp\left(-i\frac{1}{2}Tm^{2}\right) A_{m}^\nu
\end{equation}
The complex eigenvectors are localized for irrational $T$; $\abs{A_{n \rightarrow \infty}^\nu} \rightarrow 0$. The characteristic eigenvalues
$\lambda_\nu$ are complex numbers placed on the unit circle in the complex plane, $\lambda_\nu = \exp(i \chi_\nu)$. For rational $T$ then, 
extended eigenvectors are obtained.

One of the first experimentally-grounded evidences of localization in a quantum system (which leads to the suppression of the chaotic diffusion 
in the action space) was obtained with hydrogen atoms in a microwave field \cite{bayfield_localization_1989}. The quantum energy spectrum 
of this system - which consists of excited hydrogen atoms inside an intense time dependent magnetic field - is investigated in the regime when 
underlying classical motion has passed from regular to irregular behavior, via increasing magnetic field strength \cite{delande_scars_1987}. 
The first experimental realization of the quantum kicked rotor was obtained in a sample of dilute ultracold sodium atoms in a periodic standing 
wave of near-resonant light, pulsed periodically in time to approximate a series of delta kicks \cite{moore_atom_1995}. In these experiments, 
atomic momenta were measured as a function of interaction time and the pulse period. The diffusive growth of energy up to a quantum break time, 
followed by dynamical localization, was observed. In addition, in cases with the pulse period equal to an integer multiple of the rotor period, 
the ballistic diffusion and corresponding quantum resonances were observed. All these experimental findings confirmed previously established 
numerical and theoretical predictions of the quantum kicked rotor model.
\section{Nonlinear Waves: Destruction of Localization}\label{sec:nonlinear}
\subsection{Disorder}\label{sec:nonlin_dis}                   
A number of recent studies have been devoted to uncover the interplay of nonlinearity and disorder ~\cite{molina_transport_1998, pikovsky_destruction_2008,
kopidakis_absence_2008,flach_universal_2009,skokos_delocalization_2009,veksler_spreading_2009,mulansky_dynamical_2009,skokos_spreading_2010,flach_spreading_2010,
laptyeva_crossover_2010}. Most of these studies consider the evolution of an initially localized wave packet. While the linear equations will 
trap the packet, the presence of nonlinearity leads to a spreading of the packet way beyond the limits set by the linear theory. Numerical studies 
suggest that the second moment $m_2$ of the wave packet grows subdiffusively in time following a power law $t^{\alpha}$ with $\alpha < 1$. On the 
other side, for weak enough nonlinearity, wave packets appear to be frozen over the complete available integration time, thereby resembling Anderson 
localization, at least on finite time scales. As recently argued by Johansson, \etal~\cite{johansson_kam_2010}, these states may be localized for 
infinite times and correspond to Kolmogorov-Arnold-Moser (KAM) torus structures in phase space.
                      
\subsubsection{Basic models}\label{sec:basic}    
Let us consider as our first model the disordered nonlinear Schr\"{o}dinger (DNLS) chain, which has equations of motion
\begin{equation}
i \frac{\partial \psi_l}{\partial t} = \epsilon_l \psi_l - \psi_{l+1} - \psi_{l-1}  + \beta \left| \psi_l \right|^2 \psi_l 
\label{eq:DNLS}
\end{equation}
with a nonlinearity strength $\beta$ and random on-site energies chosen as in Eq.(\ref{eq:Anderson}). 

A second model we consider is the Klein-Gordon (KG) chain of coupled oscillators
\begin{equation}
\frac{\partial^2 u_l}{\partial t^2}  = -{\tilde \epsilon}_l u_l - u_l^3 + \frac{1}{W}(u_{l+1}+
u_{l-1} - 2u_l) \;,
\label{eq:KGEOM}
\end{equation}
where $\tilde{\epsilon}_l$ are uncorrelated random values chosen uniformly in the interval $[1/2, 3/2]$. 
Note in this model, $u_l$ is the generalized coordinate on the site $l$ and is wholly real, as opposed to 
Eq.(\ref{eq:DNLS})'s complex $\psi_l$ values. Nevertheless, we can reduce the linear form of 
Eq.(\ref{eq:KGEOM}) [remove the cubic $u_l^3$ term] to the same eigenvalue form as Eq.(\ref{eq:Schrodinger}).
This is done with the transforms $\epsilon_l = W(\tilde{\epsilon}_l - 1)$ and $\lambda_\nu = W\omega_\nu^2 - W - 2$, 
where $\omega_\nu$ are the KG's eigenfrequencies, $\omega_{\nu}^2 \in [1/2, 3/2 + 4/W]$. The width of the KG's 
\textbf{squared} eigenfrequency spectrum is then $\Delta=1+4/W$.

Additionally, in the KG model the total energy
\begin{eqnarray*}
E &=& \sum_l \mathcal{E}_l \\
\mbox{where}\, \mathcal{E}_l &\equiv& \frac{1}{2}(\partial u_l/\partial t)^2 + \frac{1}{2} \tilde{\epsilon}_l u_l^2 + 
\frac{1}{4} u_l^4 + \frac{1}{2W}\left( u_{l+1} - u_l \right)^2 \geq 0
\end{eqnarray*}
acts as the nonlinear control parameter, similar to $\beta$ for the DNLS case. Both models conserve the total energy; additionally, 
the DNLS conserves the total norm $\mathcal{S} = \sum_{l} |\psi_l|^2$. For small amplitudes an approximate mapping, 
$\beta \mathcal{S} \approx 3WE$, from the KG model to the DNLS model exists~\cite{kivshar_modulational_1992,kivshar_creation_1993,
johansson_discrete_2006}. Because of this mapping, in the remainder of this section we shall focus on the DNLS model's analytics,
and return to the KG model only in our observations.

The average spacing $d$ of eigenvalues of NMs within the range of a localization volume is of the order of $d \approx \Delta/V$, which becomes 
$d \approx \Delta W^2/300$ for weak disorder. The two frequency scales $d < \Delta$ are expected to determine the packet evolution details in 
the presence of nonlinearity. 

The equations of motion in Eq.(\ref{eq:DNLS}) can be rewritten in normal mode space as
\begin{equation}
i\dot{\phi}_{\nu}=\lambda_{\nu}\phi_{\nu} + \beta \sum_{\nu_1,\nu_2,\nu_3} I_{\nu,\nu_1,\nu_2,\nu_3} \phi^{\ast}_{\nu_1}
\phi_{\nu_2}\phi_{\nu_3}
\label{eq:NMEOM}
\end{equation}
where the variables $\phi_\nu = \sum_l A_l^\nu \exp(-i \lambda_\nu t)$ determine the complex time-dependent behavior of the NMs and 
$I_{\nu,\nu_1,\nu_2,\nu_3}~=~\sum_l~A_{\nu,l}A_{\nu_1,l}A_{\nu_2,l}A_{\nu_3,l}$ are the overlap integrals. The frequency shift of a single site oscillator induced by 
the nonlinearity is $\delta \sim \beta n$ for DNLS (here $n=|\psi|^2$), and $\delta \sim \mathcal{E}$ for the KG model \cite{skokos_delocalization_2009,laptyeva_crossover_2010}. 

We sort the NMs with increasing center-of-norm coordinate $X_\nu = \sum_l l (A_l^\nu)^2$. For DNLS, we monitor the time-dependent normalized 
norm density distribution in NM space, $z_\nu \equiv n_\nu / \sum_{\mu} n_\mu$. The KG counterpart is the normalized energy density distribution in NM 
space $z_{\nu} \equiv \mathcal{E}_{\nu}/\sum_{\mu}\mathcal{E}_{\mu}$. We characterize distributions by means of the second moment $m_2 = \sum_\nu 
(\nu - \bar \nu)^2 z_\nu$ (where $\bar \nu = \sum_\nu z_\nu$), which quantifies the wave packet's degree of spreading, and the participation number 
$P = 1/\sum_{\nu}z^2_{\nu}$, which measures the number of effectively excited sites. The ratio  $\zeta = P^2 / m_2$ (the compactness 
index~\cite{skokos_delocalization_2009}) quantifies the sparseness of a packet.

\subsubsection{Regimes of wave packet spreading}
We consider compact wave packets at $t=0$ spanning a width $L$ centered in the lattice, such that within $L$ there is a constant norm density of $n$ and 
a random phase at each site (outside the volume $L$ the norm density is zero). In the KG case, this corresponds to exciting each site in the width $L$ with 
the same energy density, $\mathcal{E}=E/L$, i.e. setting initial momenta to $p_l = \pm \sqrt{2\mathcal{E}}$ with randomly assigned signs. If $\delta \geq \Delta$ 
then a substantial part of the wave packet will be self-trapped ~\cite{kopidakis_absence_2008,skokos_delocalization_2009}. 
\begin{SCfigure}[10][htb]
\centering
\includegraphics[width=0.5\textwidth, keepaspectratio,clip]{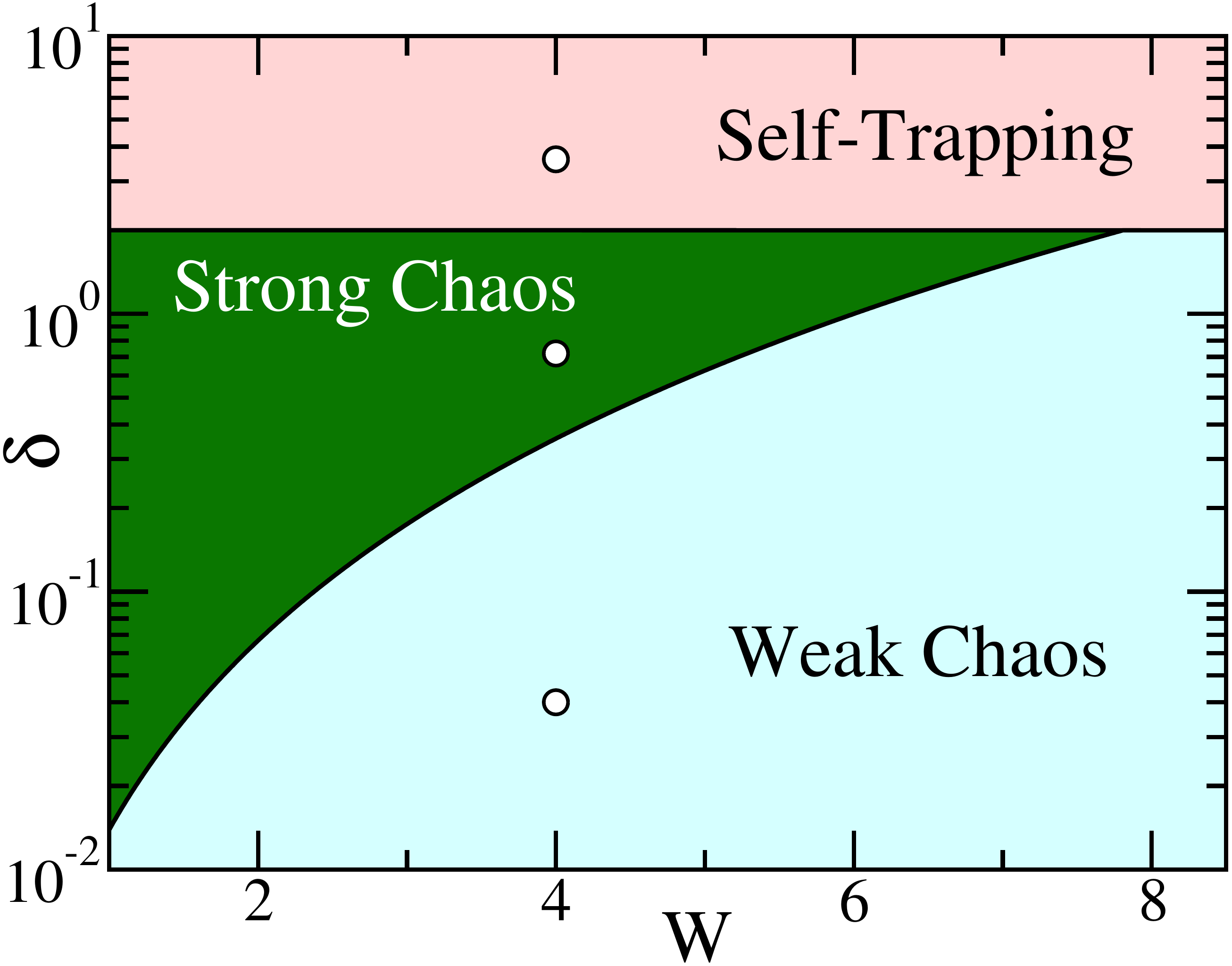}  
\caption{ Parametric space of disorder, $W$, vs. the frequency shift induced by nonlinearity, $\delta$, for the DNLS model. 
The KG analog is obtained by the small amplitude mapping. Three spreading regimes are shown for dynamics
dictated by: (i) weak chaos (pale blue), (ii) strong chaos (green), and (iii) the onset of self-trapping (pale red). 
The three circles show numerical values used in Fig.~\ref{fig:Spreading}. Adapted from \cite{laptyeva_crossover_2010}.}
\label{fig:Nonlin_Parametric}
\end{SCfigure}
This is due to nonlinear frequency shifts, which will tune the excited sites immediately out of resonance with the non-excited neighborhood ~\cite{flach_discrete_1998,
flach_discrete_2008}. In fact, partial self-trapping will occur already for $\delta \geq 2$ since at least some sites in the packet may be tuned out of resonance. 
If now $\delta < 2$, self-trapping is avoided, and the wave packet can start to spread. For $L < V$, the packet will spread over the localization volume during the 
time $\tau_{lin} \approx 2\pi/d$ (even for $\beta=0$). At that time, the new norm density will drop down to $n(\tau_{lin}) \approx (nL)/V$. For $L > V$ the norm 
density will not change appreciably up to $\tau_{lin}$ and $n(\tau_{lin}) \approx n$. The nonlinear frequency shift $\beta n(\tau_{lin})$ should be now compared 
with the average spacing $d$. If $\beta n(\tau_{lin}) > d$, all NMs in the packet are resonantly interacting with each other. We refer to this regime as strong chaos. 
If instead $\beta n(\tau_{lin}) < d$, NMs are weakly interacting with each other. We refer to this regime as weak chaos. Note that a spreading wave packet that is 
launched  in the regime of strong chaos will increase in size, drop its norm (energy) density, and therefore the crossover into the asymptotic regime of weak chaos 
must occur at later times. For a single-site excitation $L = 1$  the strong chaos regime shrinks to zero width in the norm/energy parameter and one is left only with 
either weak chaos or self-trapping ~\cite{pikovsky_destruction_2008,flach_universal_2009,skokos_delocalization_2009,veksler_spreading_2009}. 
To summarize, the expected spreading regimes for $L \geq V$ are:
\begin{equation}
\begin{aligned}
&  \delta > 2: \text{onset of self-trapping};\\
&  d < \delta < 2: \text{strong chaos};\\
&  \delta < d: \text{weak chaos}.
\end{aligned}
\label{eq:SRs}
\end{equation}
Fig.~\ref{fig:Nonlin_Parametric} sketches the predicted regimes in a parametric space for the case $L = V$, in which lines represent 
the regime boundaries $ \delta = d$ and $\delta = 2$. Note that we used $d=\Delta/(3.3 \xi(0))$ with $\xi(0)=96W^{-2}$ being the weak disorder estimate. 
\subsubsection{Resonances and chaos}
A NM with index $\mu$ in a layer of width $V$ in the cold exterior - which borders the packet but will belong to the core of the spreading packet at later 
time - is either incoherently \textit{heated} by the packet, or \textit{resonantly} excited by some particular NM from a layer with width $V$ inside the packet. 
The resonant channel will lead to spreading only if a new resonance can be found. Due to the disorder this is not possible. Then a single resonance will simply 
lead to beatings (oscillations) of the wave packet. In order to finally achieve true spreading, we have to destroy the phase coherence of the wave packet. 
Therefore there is no other way but to allow incoherent chaotic dynamics to take place, if spreading is observed.

Chaos is a combined result of resonances and nonintegrability. Let us estimate the number of resonant modes in the packet for the DNLS model. Excluding secular 
interactions, the amplitude of a NM with $\left|\phi_{\nu}\right|^2 = n_{\nu}$ is modified by a triplet of other modes $\vec{\mu} \equiv (\mu_1,\mu_2,\mu_3)$ in 
first order in $\beta$ as
\begin{equation}
\begin{aligned}
&  \left| \phi^{(1)}_{\nu} \right|=\beta \sqrt{n_{\mu_1},n_{\mu_2},n_{\mu_3}} \cdot R^{-1}_{\nu,\vec{\mu}}, \quad R_{\nu,\vec{\mu}} \sim 
\left| \frac{\vec{d \lambda}}{I_{\nu,\mu_1,\mu_2,\mu_3}} \right|,
\end{aligned}
\label{eq:PF}
\end{equation}
where $\vec{d\lambda}=\lambda_{\nu}+\lambda_{\mu_1}-\lambda_{\mu_2}-\lambda_{\mu_3}$. The perturbation approach breaks down, and resonances set in, when 
$\sqrt{n_{\nu}}<\left| \phi^{(1)}_{\nu} \right|$.  Since all considered NMs belong to the packet, we assume their norms to be equal to $n$. The main result is 
that the probability of a packet mode to be resonant is given by $\mathcal{P}= 1-e^{-C \beta n}$ \cite{krimer_statistics_2010}, with $C$ being a constant depending 
on the strength of disorder. Then
\begin{equation}
m_2 \sim \mathcal{D}\,t, \quad \mathcal{D} \sim \beta^2 n^2 \left[ \mathcal{P}(\beta n) \right]^2 
\label{eq:SM}
\end{equation}
and finally
\begin{equation}
 n^{-2} \sim \beta \left[ 1-e^{-C \beta n} \right] t^{1/2}.
\label{eq:SL}
\end{equation}
The solution of this equation yields a crossover from subdiffusive spreading in the regime of strong chaos to subdiffusive spreading in the regime of weak chaos: 
\begin{equation}
m_2 \sim 
\begin{cases}
\left[ \beta^2 t \right]^{1/2},  & C \beta n > 1  \text{ (strong chaos)}; \\
\left[ \beta^4 t \right]^{1/3}, & C \beta n  < 1 \text{ (weak chaos)}. \\
\end{cases}\label{eq:WSC}
\end{equation}
The only characteristic frequency scale here is $1/C$. From the above discussion of the different spreading regimes it follows that $1/C \approx d$. 

\subsubsection{Computational results}
\begin{SCfigure}[50][htb]
\centering
\includegraphics[width=0.6\textwidth,keepaspectratio,clip]{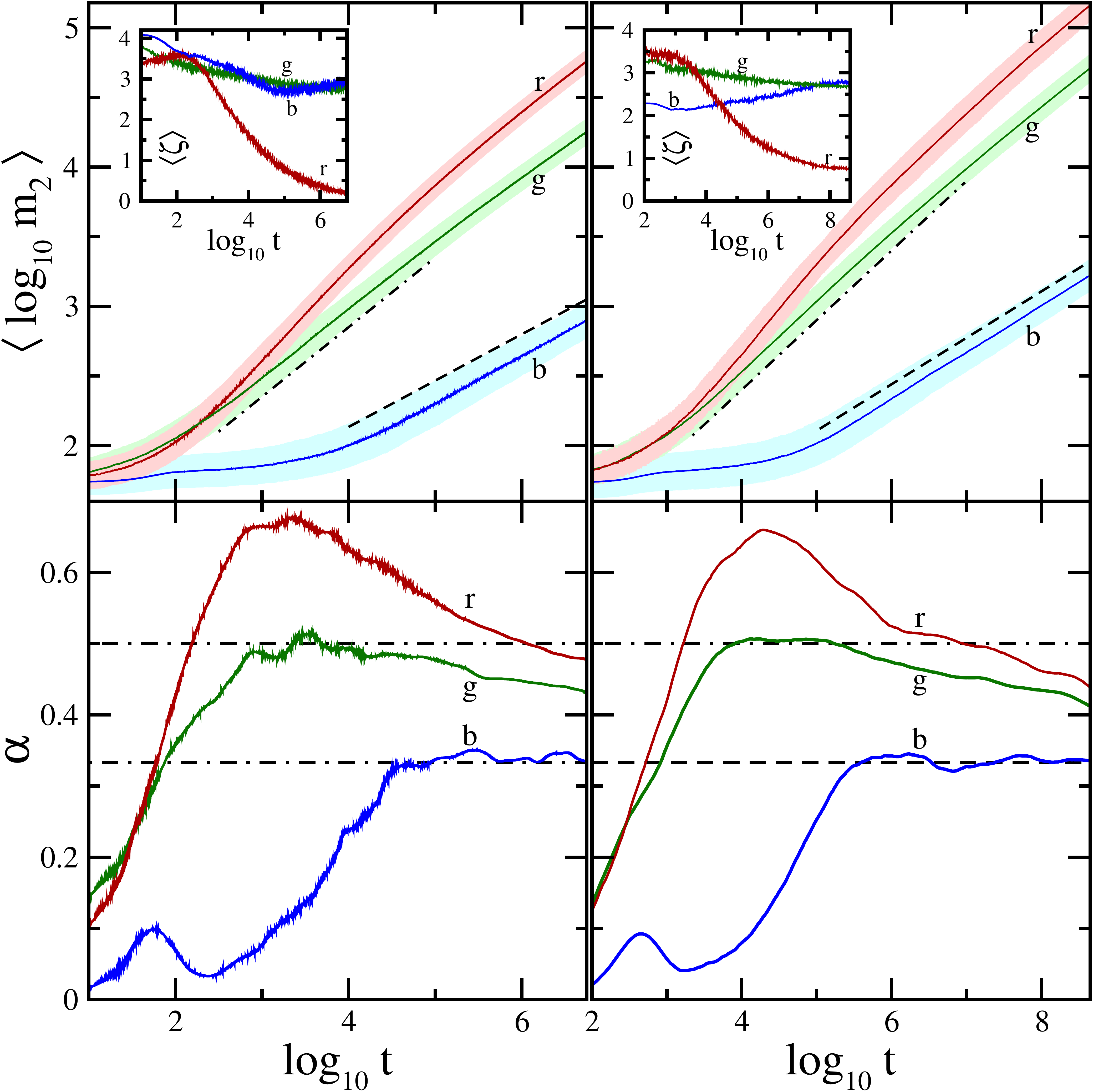}
\caption{Upper row: Average log of second moments (inset: average compactness index) vs. log time for the DNLS/KG on the  left/right, for $W=4, L=21$. 
Colors/letters correspond the three different regimes: 
(i) weak chaos - (b)lue, $\beta =0.04 \, (\mathcal E=0.01)$, 
(ii) strong chaos - (g)reen, $\beta = 0.72 \, (\mathcal E=0.2)$,
(iii) self-trapping - (r)ed, $\beta = 3.6 \, (\mathcal E=0.75)$.
The respective lighter surrounding areas show one standard deviation error. Dashed lines are to guide the eye to $\sim t^{1/3}$, 
while dotted-dashed guides for $\sim t^{1/2}$. Lower row: Finite difference derivatives $\alpha (\log t) = d \left\langle \log m_2 \right\rangle/d \log t$ 
for the smoothed $m_2$ data respectively from above curves. Adapted from \cite{laptyeva_crossover_2010}.}
\label{fig:Spreading}
\end{SCfigure}

Ensemble averages over disorder were calculated for $1000$ realizations with $W=4$ and are shown in Fig.~\ref{fig:Spreading} (upper row). We  use $L=V=21$ and system sizes 
of $1000-2000$ sites. For DNLS, an initial norm density of $n=1$ is taken, so that $\delta=\beta$. The values of $\beta$ ($\mathcal{E}$ for KG) are chosen to give 
three expected spreading regimes (see Fig.~\ref{fig:Nonlin_Parametric}), respectively  $\beta \in \left\lbrace 0.04, 0.72, 3.6\right\rbrace$ and $\mathcal{E} \in \left\lbrace 0.01, 
0.2, 0.75\right\rbrace$. In the predicted regime of weak chaos we indeed find a subdiffusive growth of $m_2$  according to $m_2 \sim t^\alpha$ with $\alpha \approx 1/3$ 
at large times. In the expected regime of strong chaos we observe exponents $\alpha\approx 1/2$ for $10^3 \lesssim t \lesssim 10^4$ (KG: $10^4 \lesssim t \lesssim 10^5$) 
in Fig.~\ref{fig:Spreading}. Time averages in these regions over the green curves yield $\alpha \approx 0.49\pm 0.01 \;(\mbox{KG: } 0.51\pm 0.02)$. With spreading continuing in the 
strong chaos regime, the norm density in the packet decreases, and eventually satisfies $\beta n \leq d$. That leads to a dynamical crossover to the slower weak chaos 
subdiffusive spreading, as predicted. Fits of the further decay suggest $\alpha \approx 1/3$ at $10^{10} \lesssim t \lesssim 10^{11}$. 
\begin{SCfigure}[50][htb]
\centering
\includegraphics[width=0.6\textwidth,keepaspectratio,clip]{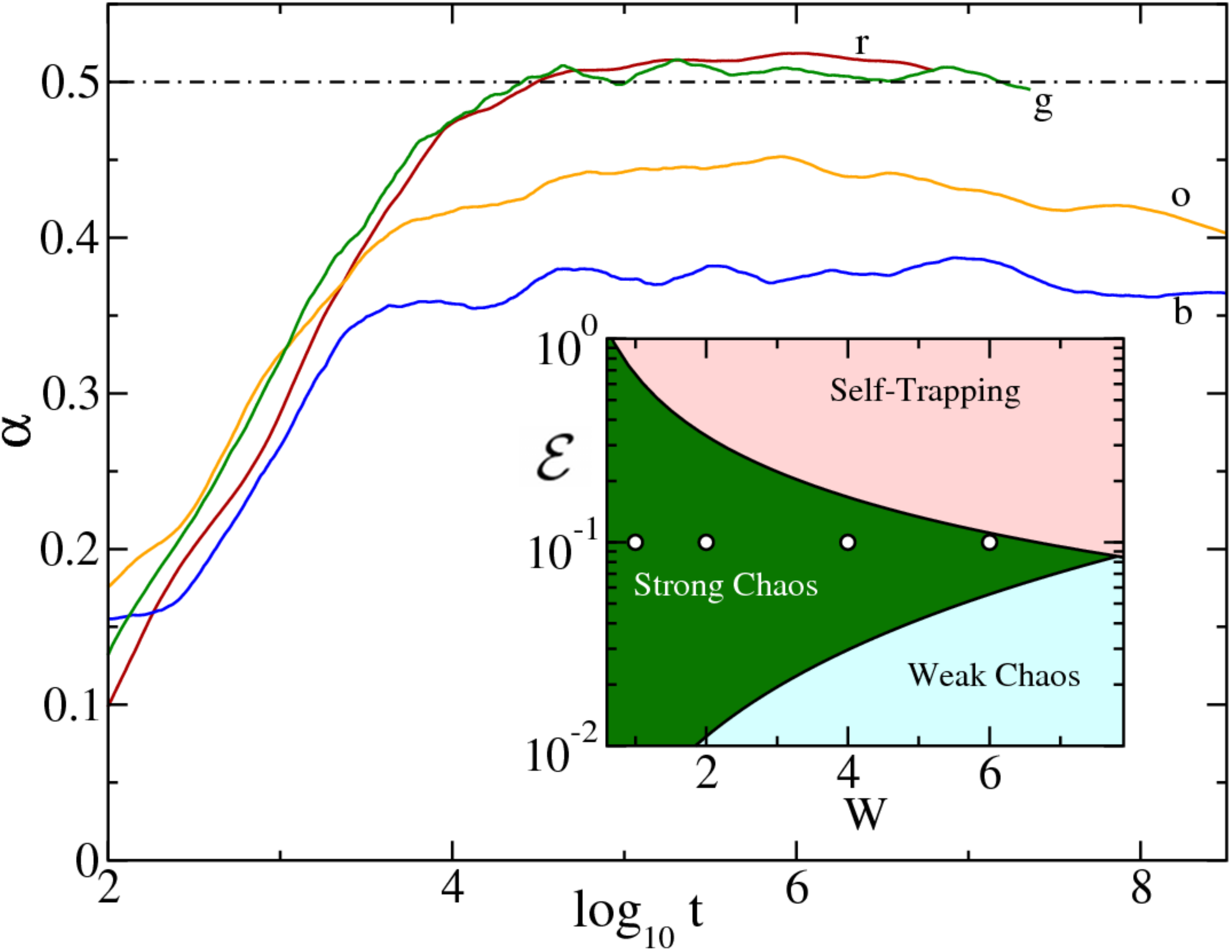}
\caption{Spreading behavior in the strong chaos regime for the KG model, with an initial energy density of $\mathcal{E}=0.1$. The four curves are for the 
disorder strengths of: $W=1$ - (r)ed, $W=2$ - (g)reen, $W=4$ - (o)range, $W=6$ - (b)lue. Inset: the KG analog of the DNLS parametric space from Fig.~\ref{fig:Nonlin_Parametric}. 
The four points correspond to the disorder strengths used in the main portion of the figure. Adapted from \cite{laptyeva_crossover_2010}.}
\label{fig:Strong_KG}
\end{SCfigure}
In the regimes of weak and strong chaos, the compactness index at largest computational times is $\zeta \approx 2.85 \pm 0.79 \; (\mbox{KG: }2.74\pm0.83)$, as 
seen in the blue and green curves of Fig.~\ref{fig:Spreading}. This means that the wave packet spreads, but remains rather compact and thermalized ($\zeta \approx 3$).
The duration of $\alpha = 1/2$ spreading (and the crossover time) is largely dependent on how deep we are initially in the strong chaos regime. 
This is illustrated in Fig.~\ref{fig:Strong_KG} for the KG model. For $W \in \left\lbrace 1,2\right\rbrace $ a long-lasting strong chaos spreading is clearly observed. 
For $W \in \left\lbrace 4,6 \right\rbrace$ the width in the energy density is small and, even if, initially, the energy density is chosen to give strong chaos, 
its decrease due to spreading will get the system into the weak chaos regime with $\alpha < 1/2$. 

\subsubsection{Generalizations}
Let us apply the same theoretical arguments to a general $\mathbf{D}$-dimensional lattice with the nonlinearity of the  order $\sigma$:
\begin{equation}
i \dot{\psi_{\mathbf{l}}} = \epsilon_{\mathbf{l}} \psi_{\mathbf{l}} - \beta \left| \psi_{\mathbf{l}} \right|^{\sigma} \psi_{\mathbf{l}} - 
\sum \limits_{{\mathbf{m}} \in \mathbf{D}({\mathbf{l}})} \psi_{\mathbf{m}}.
\label{eq:gDNLS}
\end{equation}
Here ${\mathbf{l}}$ denotes a $\mathbf{D}$-dimensional lattice vector with integer components, and ${\mathbf{m}} \in \mathbf{D}({\mathbf{l}})$ 
defines its set of nearest neighbors. We assume that all NMs are spatially localized (which can be obtained for strong enough disorder W). 
A wave packet with the same average norm $n$ per excited mode has a second moment $m_2 \sim n^{-2/\mathbf{D}}$. The nonlinear frequency shift 
is proportional to $\beta n^{\sigma/2}$.

A straightforward generalization of the expected regimes of wave packet spreading \cite{flach_spreading_2010} with $L \geq V$ leads to the 
following: self-trapping if $\beta n^{\sigma/2}>\Delta$, strong chaos if $\beta n^{\sigma/2}>d$, and weak chaos if $\beta n^{\sigma/2}<d$. 
The regime of strong chaos can be observed for $n > \left[ d/\beta \right]^{2/\sigma}$.

Similar to the above we obtain a diffusion coefficient
\begin{equation}
\mathcal{D} \sim \beta^{2} n^{\sigma} \left[ \mathcal{P}(\beta n^{\sigma/2}) \right]^2.
\label{eq:gDK}
\end{equation}
In both regimes of strong and weak chaos the spreading is subdiffusive \cite{flach_spreading_2010}:
\begin{equation}
m_2 \sim 
\begin{cases}
\left[ \beta^2 t \right]^{\frac{2}{2+\sigma \mathbf{D}}}, \text{ (strong chaos)}; \\
\left[ \beta^4 t \right]^{\frac{1}{1+\sigma \mathbf{D}}}, \text{ (weak chaos)}. \\
\end{cases}\label{eq:gWSC}
\end{equation}

\begin{SCfigure}[50][h]
\centering
\includegraphics[width=0.6\textwidth,keepaspectratio,clip]{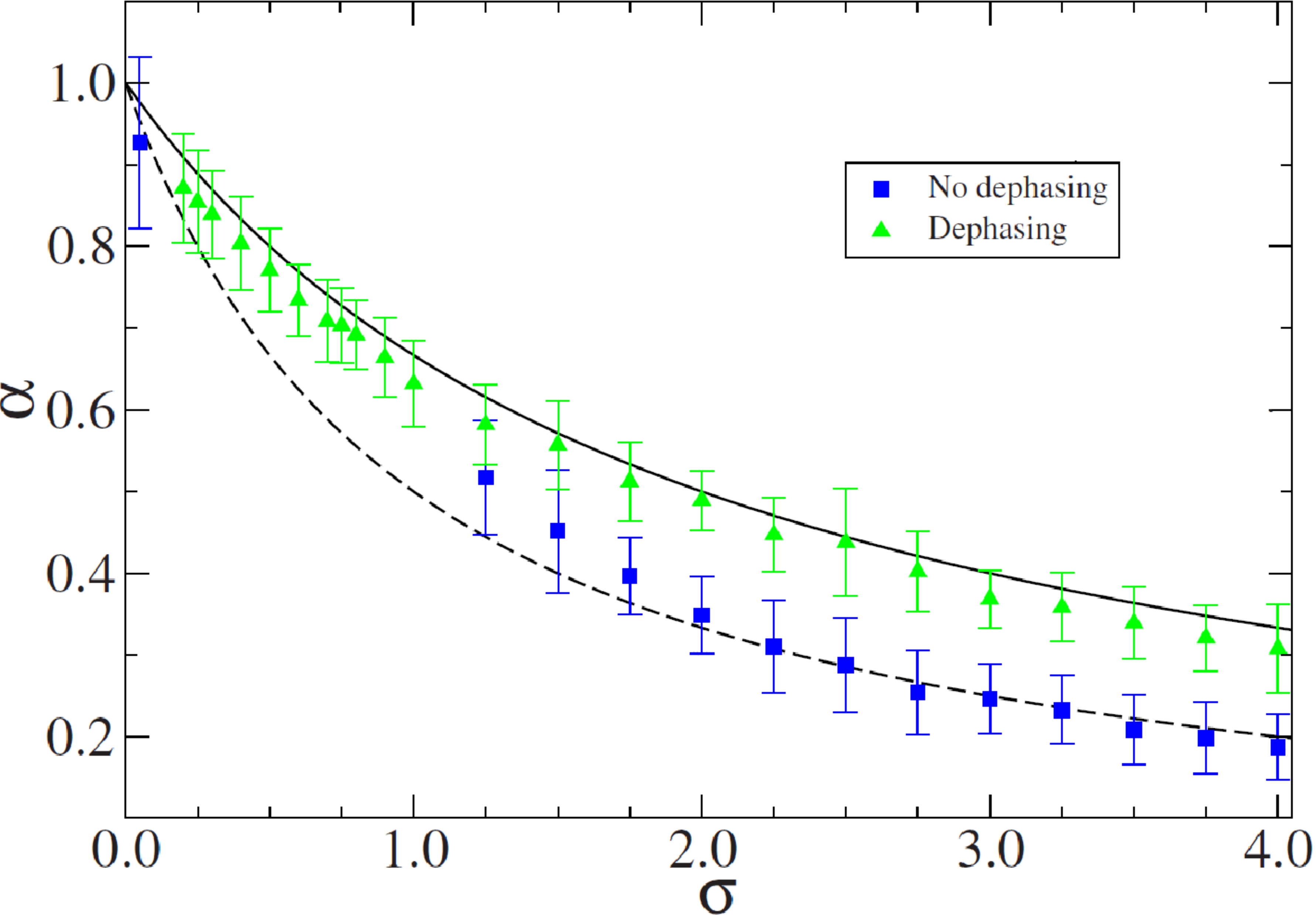}
\caption{Spreading exponent $\alpha$ vs. the nonlinearity power $\sigma$ for integration without dephasing (filled squares) and for integration 
with dephasing of NMs (filled triangles). The theoretically predicted boundaries of weak chaos and strong chaos are plotted by dashed and solid 
lines, respectively. Adapted from \cite{skokos_spreading_2010}.}
\label{fig:Nonlin_Tuning}
\end{SCfigure}
The number of resonances on the wave packet surface is $N_{RS} \sim \beta n^{\frac{\mathbf{D}(\sigma-2)+2}{2 \mathbf{D}}}$. 
This number increases with time for 
\begin{equation}
\mathbf{D}>\mathbf{D}_c=\dfrac{1}{1-\sigma/2}, \; \sigma<2.
\label{eq:gDim}
\end{equation}

We expect that the wave packet surface will not stay compact if Eq.(\ref{eq:gDim}) is fulfilled \cite{flach_spreading_2010}. Instead, surface resonances 
will lead to a resonant leakage of excitations into the exterior. This process will increase the surface area, and therefore lead to even more surface 
resonances, which increase the surface area further on. The wave packet will fragmentize, perhaps get a fractal-like structure, and lower its compactness 
index. The spreading of the wave packet will speed up, but will not anymore be due to pure incoherent transfer, instead it will become a complicated 
mixture of incoherent and coherent transfer processes. For such cases, Anderson localization will be destroyed quickly even in the tails of wave 
packets \cite{skokos_spreading_2010}.

The numerical evidence \cite{skokos_spreading_2010} for the validity of predictions Eq.(\ref{eq:gWSC}) for the generalized KG model 
$\ddot{u}_l  = -{\tilde \epsilon}_l  u_l - \left|u_l\right|^{\sigma}u_l + \frac{1}{W}(u_{l+1}+u_{l-1} - 2u_l)$ 
and energies away from the self-trapping regime are presented in Fig.~\ref{fig:Nonlin_Tuning}. The energy values used there cross the boundary 
between the weak and strong chaos regimes around the interval $1 \lesssim \sigma \lesssim 2$. In particular, the computed exponents are in a 
good agreement with the theoretical prediction for weak chaos Eq.(\ref{eq:gWSC}) for $\sigma \geq 2$. For smaller values of $\sigma$ they smoothly 
cross over to the prediction of strong chaos, as expected.

\subsection{Wannier-Stark Ladder}\label{sec:nonlin_stark}
The evolution of a wave packet in a nonlinear Wannier-Stark ladder was studied in \cite{krimer_delocalization_2009}. Nonlinearity induces frequency shifts and 
mode-mode interactions and destroys localization. For large strength of nonlinearity we observe single-site trapping as a transient, with subsequent 
explosive spreading, followed by subdiffusion. For moderate nonlinearities an immediate subdiffusion takes place. Finally, for small nonlinearities 
we find linear Wannier-Stark localization as a transient, with subsequent subdiffusion. For single-mode excitations additional stability and instability 
intervals with respect to the DC bias strength were shown to exist. The onset of subdiffusive spreading was also observed in 
Refs.~\cite{datta_effect_1998,kolovsky_bose-einstein_2010} for two runs on rather short scales up to $t=10^5$.

\subsubsection{Basic model}
We consider the discrete nonlinear Schr\"{o}dinger equation with a DC bias $E$
\begin{equation}
i{\partial \psi_l}{\partial t} = l E \psi_l - \psi_{l+1} - \psi_{l-1} + \beta|\psi_l|^2\psi_l,
\label{eq:Nonlin_Stark}
\end{equation}
As in Eq.(\ref{eq:NMEOM}), a transformation to the NM space first is made using $\phi_\nu = \sum_l A_l^\nu \exp(-i \lambda_\nu t)$. 
The linear term can be gauged out by use of a secular normal form $\phi_\nu = \chi_\nu \exp(-i \nu E t)$, yielding
\begin{equation}
i\frac{\partial \chi_\nu}{\partial t}=\beta\!\!\sum\limits_{\nu_1,\nu_2,\nu_3}I_{\nu,\nu_1,\nu_2,\nu_3}\chi_{\nu_1}^*\chi_{\nu_2} \chi_{\nu_3}e^{i(\nu+\nu_1-\nu_2-\nu_3)\,Et}.
\label{eq:NM_Stark}
\end{equation}
where
\begin{eqnarray}
I_{\nu,\nu_1,\nu_2,\nu_3}\equiv\sum\limits_{n} A_{n-\nu}^{(0)} A_{n-\nu_1}^{(0)} A_{n-\nu_2}^{(0)} A_{n-\nu_3}^{(0)}
\label{eq_overl_Stark}
\end{eqnarray}
are the overlap integrals between the NMs. As for DNLS with disorder, adding nonlinearity again leads to a finite range interaction between the 
eigenstates. The main difference from the disordered case is that here, the linear spectrum is unbounded and exact resonances are always present. 
Resonant normal form equations are indeed obtained by substituting $\nu+\nu_1-\nu_2-\nu_3=0$ into Eq.~(\ref{eq:NM_Stark}). These equations are not 
integrable \cite{krimer_delocalization_2009}, in contrast to the resonant normal form equations for the disordered case \cite{flach_spreading_2010}. 
As a consequence, if at least two neighboring NMs are excited, the resonant normal form, having a connectivity similar to the original lattice 
equations, allows spreading over the whole lattice. Such a direct resonant interaction mechanism takes place between the NMs both inside and outside the 
wave packet.

\subsubsection{Single site excitation}
\begin{figure}[htb]
\centering
\includegraphics[width=\textwidth, keepaspectratio,clip]{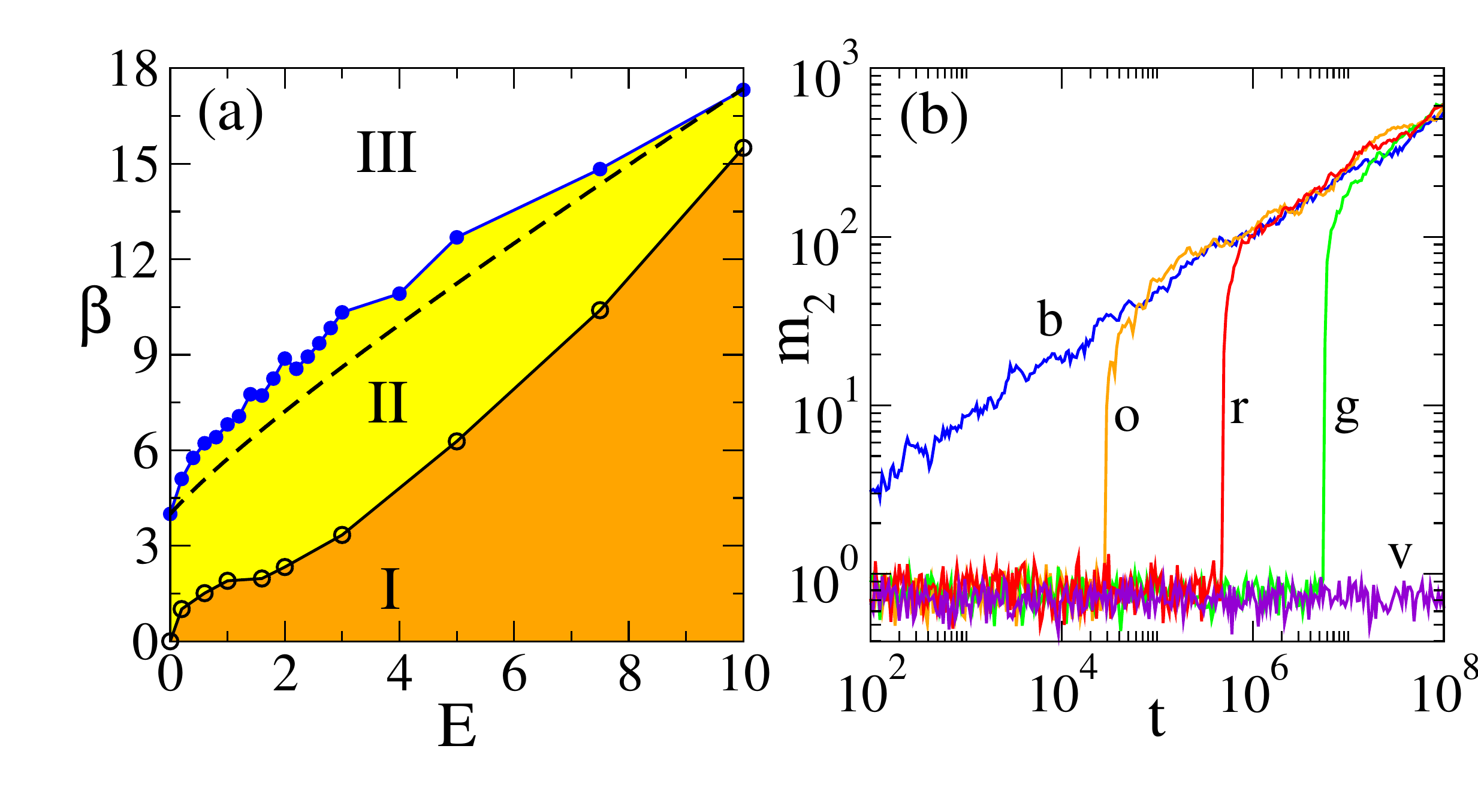}  
\caption{(a) The diagram of the three regimes of spreading in the parameter space $(\beta,E)$ for Eq.(\ref{eq:Nonlin_Stark}). Empty and filled circles: numerically 
obtained thresholds which separate the three different regimes I, II, III - lines connecting symbols are meant to guide the eye for the boundaries. 
Black dashed line: threshold between II and III obtained from the dimer model. In the limit of $E\rightarrow\infty$, all lines merge to the asymptotic limit 
$\beta \propto E$. (b) Single site excitation for $E=2$. Second moment $m_2$ versus time in log-log plots for different values of $\beta$ inside the interval 
where an explosive delocalization of the trapped regime occurs: $\beta=8.15,8.25,8.5$  [(o) orange; (g) green; (r) red]. $\beta=8$ [(b), blue]: intermediate regime. 
$\beta=8.9$ [(v), violet]: trapped regime. Figure adapted from \cite{krimer_delocalization_2009}. }
\label{fig:Nonlin_Stark}
\end{figure}

First, we study a single site initial excitation $\psi_l(0)=\delta_{l0}$. In that case the amplitudes in NM space are $\phi_\nu(0)=J_\nu(2/E)$. 
The nonlinear frequency shift, $\delta \sim \beta$, at site $n=0$ should be compared with the two scales set by the linear problem: the eigenvalue 
spacing $E$ and the eigenvalue variation over a localization volume $\Delta \equiv E\mathcal{L}$ (see Sec.~\ref{sec:lin_stark}).  We found three qualitatively 
different regimes of spreading shown on the phase diagram in the parameter plane of nonlinearity, $\beta$, and DC field, $E$,  strengths 
(see Fig.~\ref{fig:Nonlin_Stark}a): (I) $\delta < E$, (II) $E < \delta < \Delta$, (III) $ \Delta < \delta$. In case (I), the nonlinear frequency shift is less 
than the spacing between excited modes. Therefore no initial resonance overlap is expected, and the dynamics may evolve as the one for $\beta=0$ at least 
for long times. In case (II), resonance overlap happens, and the packet expands subdiffusively from the very beginning. For case (III), $\delta$ tunes 
the excited site out of resonance with the neighboring NMs. Resonances with more distant NMs are possible, but the overlap with these NMs is the weaker 
the further away they are. Therefore for long times the excited site may evolve as an independent oscillator (trapping). 

Let us discuss case (III) in more details. For $E=2$ and $\beta>8.9$, the single site excitation stays trapped up to times $t=10^8$ without significantly spreading 
into any other site of the lattice (violet curve in Fig.~\ref{fig:Nonlin_Stark}). Slightly lowering $\beta$ we observe that the excitation is trapped up to a some time $T_E$ 
which sensitively depends on $\beta$ and changes by many orders of magnitude e.g. between $10^2$ to $10^7$ in the narrow interval $\beta\in (8.05,8.9)$ for $E=2$
(Fig.~\ref{fig:Nonlin_Stark}). For times $t > T_E$ an explosive and spatially asymmetric spreading is observed on a time scale of one Bloch period $T_B$. The packet spreads 
in the direction of NMs with larger eigenvalues, which provide the possibility of resonant energy transfer from the single site excitation due to its positive nonlinear 
frequency shift $\delta$. For about ten Bloch periods $T_B$ the packet shows  Bloch oscillations, which then quickly decohere. Finally the packet spreads incoherently 
and subdiffusively. The explosion time $T_E$ is not monotonously changing with $\beta$, which indicates intermittency, \ie the single site excitation can be closer or 
further away from some regular structures in phase space. That distance may in turn control the value of $T_E$. For $E=2$ and $\beta=8$ 
the packet spreads from scratch (blue curve in Fig.~\ref{fig:Nonlin_Stark}). 

It is worth noting, that the border between regime II and III can be approximated by a dimer model (an estimate from below). Indeed, a dimer model takes into account 
only one lattice site to the right from initially excited site (corresponding to the trapped state) and therefore describes the asymmetric energy transfer during the explosion.
As for the disordered case, we observe that nonlinearity destroys integrability, introduces chaos, and ultimately leads to a subdiffusive spreading, so that the second moment 
grows as $t^\alpha$ with $\alpha<1$. Our preliminary numerical studies showed that the exponent $\alpha$ is not universal and depends on the system parameters. The reason is 
that spreading of the wave packet takes place not anymore due to pure incoherent transfer but becomes a complicated mixture of incoherent and coherent transfer processes. 
The interplay of these two mechanisms is a subject of our future studies.

\subsubsection{Single mode excitation}
A single mode excitation $\phi_\nu(t=0)=\delta_{\nu,0}$ also exhibits the three different regimes of spreading. However, for small values of nonlinearity $\beta$ a new 
intriguing feature of the short time dynamics follows \cite{krimer_delocalization_2009}. Indeed, considering the resonant normal form one can conclude that a single mode 
excitation is the exact solution, so that no other NM is going to be excited. However, the full set of equations (\ref{eq:NM_Stark}) will excite other NMs as well. 
These small perturbations may stay small or start to grow, depending on the stability of the single mode solution. Performing the linear stability analysis of the 
single mode excitation the stability intervals, which affect the short and long time dynamics, are obtained and observed upon variation of the DC bias strength. 
In particular, if the single mode excitation is launched within a stability window, we found that the wave packet is practically not spreading up to long time scales. 
However, a small change of the DC bias value $E$ tunes the system into an instability window leading to a subdiffusive spreading on the accessible time scales, starting 
at short time scales.
\subsection{Nonlinear Aubry-Andr\'{e} Chains}
In quasiperiodic systems, the localized-delocalized transition discussed in  Section~\ref{sec:quasi} may be probed by a nonlinear interaction.
Just as in Eqs.(\ref{eq:DNLS},\ref{eq:Nonlin_Stark}), this is done by addition of a cubic term in the dynamics of Eq.(\ref{eq:harper}):
\begin{equation}
i \frac{\partial \psi_l}{\partial t} =  \zeta \cos(2 \pi \alpha l) \cdot \psi_l -  \psi_{l+1} - \psi_{l-1} + \beta \abs{\psi_l}^2 \psi_l \label{eq:nonlin_harper}
\end{equation}

\begin{SCfigure}[50][h]
\centering
\includegraphics[width=0.5\textwidth,keepaspectratio,clip]{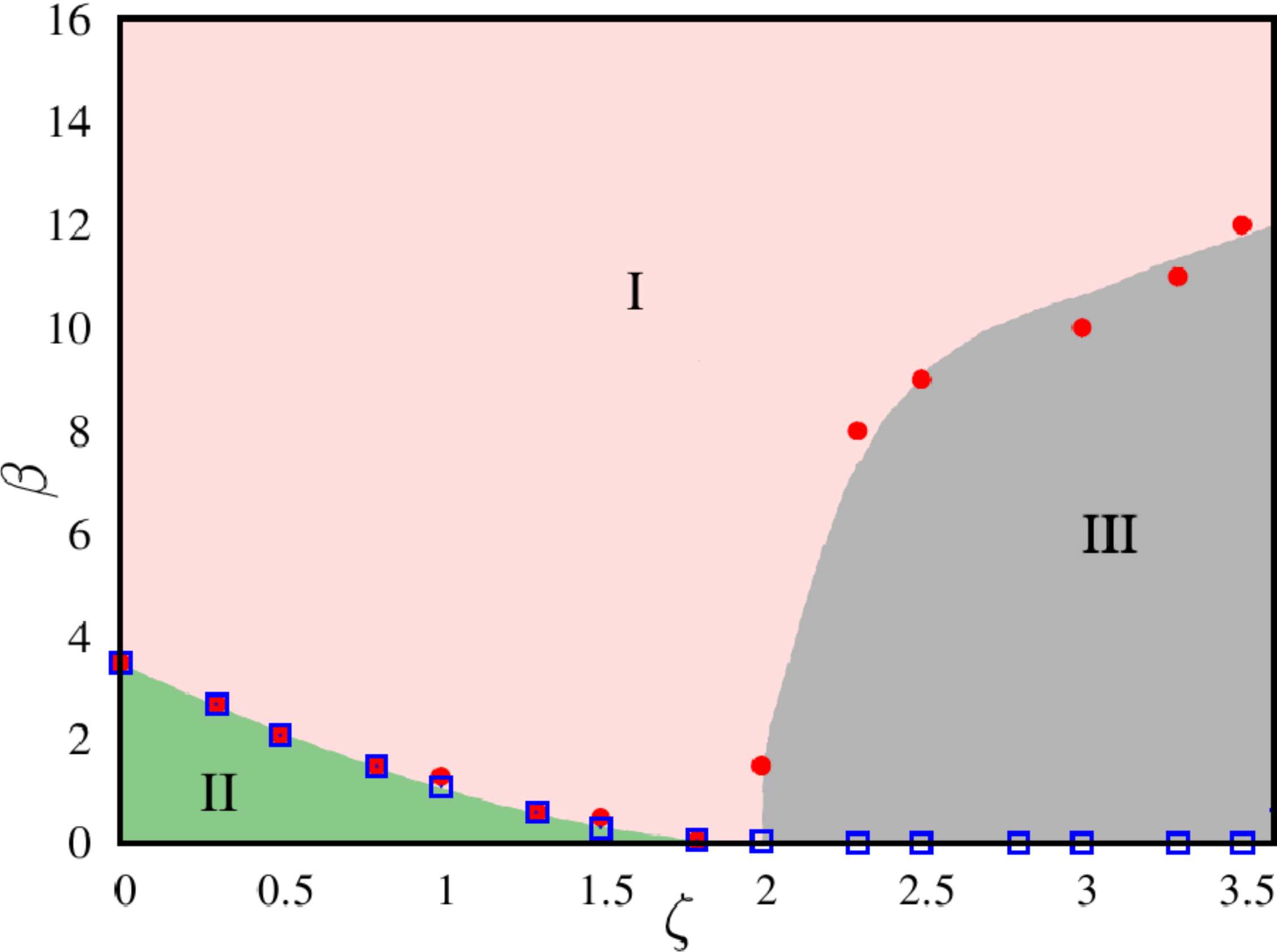}
\caption{Parametric space for nonlinear spreading in the Aubry-Andr\'{e} model. The blue squares are for an initial excitation with zero phase, 
while the red circles are for an excitation of $\pi$ phase. The three regimes are discussed in detail within the text. Figure adapted from 
\cite{larcher_effects_2009}.}
\label{fig:Nonlin_Incomm}
\end{SCfigure}

Unlike the two previous discussed cases, the \textbf{linear} behavior of the second moment (introduced in Sec.~\ref{sec:basic}) in the Aubry-Andr\'{e} 
model was shown \cite{ketzmerick_what_1997,hiramoto_dynamics_1988} to follow
\begin{equation*}
m_2(t; \beta=0) =  
\begin{cases}
   t^1 &  \zeta < 2 \\
   t^{1/2} & \zeta = 2 \\
   t^0 & \zeta > 2
  \end{cases}
\end{equation*}
The nonlinear effect on the above packet spreading has garnered much attention in recent years, including experimental observations of the spreading 
in both Kerr photonics \cite{lahini_observation_2009} and ultracold atomic clouds in optic traps \cite{ringot_experimental_2000, 
deissler_delocalization_2010}. In \cite{ng_wavepacket_2007}, the critical case of $\zeta=2$ was observed to have short transient behaviors 
dependent on the nonlinearity before asymptotically displaying an exponent similar to the linear behavior. In contrast, in 
\cite{larcher_effects_2009} all $\zeta$ were investigated. Starting from single-site excitations, they were able to develop the parametric space 
shown in Fig.~\ref{fig:Nonlin_Incomm}.
In this figure, three spreading regimes are found to be dictated by the initial condition of the single-site excitation. The blue square
datapoints in the figure correspond to an initial condition of a zero phase, while the red circles correspond to a $\pi$ phase. The first 
two regions display (I) strong self-trapping (similar to the $\delta > 2$ section of Fig.~\ref{fig:Nonlin_Parametric}), and (II) subdiffusive 
spreading, but with discrete breather structures being seen. The main interest is within the localizing region (III), in which initial zero 
phases become self-trapped, and initial $\pi$ phases become subdiffusive - these dependencies were hinted within \cite{johansson_effects_1995}. 
Larcher \etal then go on to investigate the exponents in the moments and participations, much as done for the DNLS and KG models in 
\cite{flach_universal_2009}. Throughout, the spreading still remains subdiffusive. Recent efforts have also seen experimental evidence
to support such subdiffusive spreading \cite{lucioni_observation_2010}.

Almost all efforts in the Aubry-Andr\'{e} model have been done for a single 'realization' of the potential. However, to develop a universal 
descriptions of spreading, moment averaging over realizations ought be considered - while there is no randomness in Eq.(\ref{eq:nonlin_harper}), 
one can introduce the idea of a 'realization' by a random lattice phase $\theta:\, \cos(2 \pi \alpha l)\mapsto \cos(2 \pi \alpha l + \theta)$ or 
perhaps over various incommensurate sets of $\alpha$. As seen, particular attention needs paid in choosing the initial condition. With the 
introduction of nonlinearity, the Aubry-Andr\'{e} model remains a cornucopia of study. 
\subsection{Quantum Kicked Rotor}
Realizations of Bose-Einstein condensation of dilute gases has opened new opportunities for experimental study of dynamical systems in the presence 
of many-body interactions. In the mean field approximation, these many-body interactions in the Bose-Einstein condensates are represented by adding 
a nonlinear term in the corresponding Schr\"odinger model equation \cite{benvenuto_manifestations_1991}:
\begin{equation}
i\frac{\partial \psi}{\partial t} = -\frac{1}{2}\frac{\partial^{2} \psi}{\partial \theta^{2}}-\tilde{\beta}\left|\psi\right|^{2} \psi + 
k cos\left(\theta\right)\psi \sum^{+\infty}_{m=-\infty} \delta \left(t-mT\right) \,\,,\label{this0}
\end{equation}
where the notation is the same as Eq. (\ref{this01}), except here with the new parameter $\tilde{\beta}$, which describes the nonlinear coupling.

The influence of nonlinearity on quantum localization in the nonlinear quantum kicked rotor can be studied by direct numerical simulation of the 
corresponding model, Eq.(\ref{this0}). The correct approach to approximate the evolution of the nonlinear Schr\"{o}dinger equation is to evaluate 
the nonlinear term in the position representation \cite{benvenuto_manifestations_1991}. Namely, for the numerical integration of Eq.(\ref{this0}),
the lowest order split method can be used and the evolution operator $\hat{U}$ can be approximated by the time-ordered product of the evolution
operators over small time steps $T/L$ (with integer $L$) \cite{bandrauk_exponential_1993}:
\begin{equation}
\hat{U}\left(T\right)=\exp\left(-ik\cos\hat{\theta}\right)\prod^{L}_{j=1}\exp\left(-iT\frac{\hat{n}^{2}}{2L}\right)\exp\left(i\tilde{\beta}
\left(T/L\right)\left|\psi\left(\hat{\theta},jT/L\right)\right|^{2}\right)\,\,.\label{this6}
\end{equation}
In this model the phase is acquired at each instant by the wavefunction, which involves all the Fourier components, and the phase factor of the $n$th 
Fourier component is $\left(\tilde{\beta}/2\pi\right)\sum_{m}\hat{\psi}^{*}_{m+n}\hat{\psi}_{m}$. The typical values of the number of steps per period 
are between $8\cdot 10^5$ and $5\cdot 10^6$. Therefore, this model is computationally quite expensive to study the effects of strong nonlinearities and
the dynamics of the system over long time \cite{rebuzzini_delocalized_2005}.

Another much simpler model of the quantum kicked nonlinear rotor - which allows faster performance and more efficient numerical
computations - was introduced by Shepelyansky \cite{shepelyansky_delocalization_1993}. The dynamics of this model is given by
the following map:
\begin{equation}
A_{n}\left(t+T\right)=\sum_{m}\left(-i\right)^{n-m}J_{n-m}\left(k\right)A_{m}\left(t\right)\exp\left(-i\frac{1}{2}Tm^{2}+
i\beta^{~}\left|A_{m}\right|^{2}\right)\,\,,\label{this4}
\end{equation}
This map is almost the same as that without nonlinearity, Eq.(\ref{this31}). The only difference is that the change of the phase in the Fourier harmonics 
$A_{n}$ between two kicks, which now depends on the amplitude of the harmonics, $\Delta\phi_{m}=\beta\left|A_{m}\right|^{2}$. The parameter of the nonlinear
coupling $\tilde{\beta}$ in the nonlinear Schr\"{o}dinger equation Eq.(\ref{this0}) and the nonlinear parameter $\beta$ in Eq.(\ref{this4}) are 
connected by the relation $\beta=T\tilde{\beta}/2\pi$.

Numerical results of the quantum kicked rotor model in the presence of nonlinearity show that the dynamics is affected by nonlinearity. In the resonant 
regime, where the parameter $T$ is rational, the nonlinearity only affects the prefactor in the parabolic growth law, and the resonant regime persists
\cite{rebuzzini_delocalized_2005}. On the other hand, in the localized regime (irrational $T$) simulations demonstrate that nonlinearity destroys quantum 
localization. Namely, in the presence of strong enough nonlinearity, subdiffusive spreading is observed \cite{shepelyansky_delocalization_1993}. This effect 
of nonlinearity in the quantum kicked rotor is similar to those obtained in the models with disorder, as discussed in the previous sections. In the case of 
the quantum kicked rotor, the role of disorder is played by the quasiperiodic sequence $\left\{\frac{1}{2} T m^{2}\right\}$, which is obtained for irrational 
$T$. Replacement of this quasiperiodic sequence with a truly random one shows no change in the behavior of the system. This favors the expectations that the 
influence of nonlinearity can be described by the same theory developed in the context of the models with disorder. Preliminary results show that the 
compactness index $\zeta$ (see Sec.~\ref{sec:basic}), which quantifies the wave packet sparsity, oscillates around values of $12$ for the quantum kicked 
rotor model. This means that the wave packet spreads in a compact fashion. In addition, preliminary results indicate different subdiffusive spreading 
regimes with respect to the values of the coefficient of nonlinearity $\beta$ and the strength of the kick $k$. We expect that these results can be explained 
by the existence of different spreading regimes of the wave packet, as in Sec.~\ref{sec:nonlin_dis}: the strong chaos, the weak chaos regime, and the crossover 
regime between. Currently, our recent efforts have been dedicated to establishing more clear and reliable results.
\section{Outlook}
A variety of linear wave equations support wave localization. Many of them were experimentally studied in recent times. The localization phenomenon 
relies on the phase coherence of waves. Therefore, destruction of phase coherence - dephasing - leads to a loss of wave localization. Nonlinear wave 
equations are in general nonintegrable, and therefore admit dynamical chaos. Dynamical chaos in turn leads to a loss of correlations - and therefore
dephasing. Consequently wave propagation in nonlinear wave equations will generically lead to delocalization. We discussed a number of results on wave 
packet propagation which support this conclusion. Nonlinearity is the result of wave interactions, and a generic phenomenon in many physical realizations 
as well. Therefore, future experimental studies are expected to confirm these predictions.

Besides wave packet propagation, conductivity measurements are informative as well. In particular, the temperature dependence of the heat conductivity 
has been recently related directly to the properties of wave packet propagation \cite{flach_thermal_2011}.

There are many future research directions the mind can take. Extensions to higher lattice dimensions are of interest. Two-dimensional lattices are
numerically feasible, while three-dimensional cases need most probably the support of supercomputers. The interplay of nonlinearity with mobility edges 
and critical states can be expected to be intriguing as well. Quantizing the above nonlinear field equations remains a challenging enterprise. 

\nonumsection{Acknowledgments} \noindent
The authors wish to acknowledge several fruitful discussions with S. Aubry, M.V. Ivanchenko, T. Bountis, G. Modugno, M. Modugno, M. Inguscio, Y. Silberberg, 
M. Segev, R. Schilling, S. Fishman, A. Soffer, B.L. Altshuler, A. Pikovsky, D. Shepelyanksy, V. Oganesyan, I. Aleiner, D. Basko, R. Khomeriki, N. Li and W. Wang.
Ch.~Skokos was partly supported by the European research project ``Complex Matter'', funded by the GSRT of the Ministry Education of Greece under the 
ERA-Network Complexity Program.

\bibliographystyle{ws-ijbc}

\end{document}